\RequirePackage{fix-cm}
\documentclass{svjour3}                     
\smartqed  
\usepackage{graphicx}
\usepackage{epstopdf}
\usepackage{aas_macros}
\usepackage{subfigure}
\usepackage{mathtools}
\usepackage{caption}
\usepackage{natbib}
\usepackage{color}

\begin{document}

\title{Stability and Fourier-series periodic solution in the binary stellar systems
}

\titlerunning{Stability and Fourier-series periodic solution in the binary stellar systems}        

\author{Rajib Mia         \and
        Badam Singh Kushvah 
}


\institute{R. Mia \at
              Department of Applied Mathematics, Indian School of Mines, Dhanbad 826004, Jharkhand, India 
              \\
              Tel.: +91-3262235765\\
              Fax: +91-3262235765\\
              \email{rajibmia.90@gmail.com}
           \and
           B.S. Kushvah \at
           Department of Applied Mathematics, Indian School of Mines, Dhanbad 826004, Jharkhand, India\\
             \email{bskush@gmail.com}
}

\date{Received: date / Accepted: date}

\maketitle

\begin{abstract}
In this paper, we use the restricted three body problem in the binary stellar systems, taking photogravitational effects of both the stars. The aim of this
study is to investigate the motion of the infinitesimal mass in the vicinity of the Lagrangian points. We have computed semi-analytical expressions for the locations of the collinear points with the help of the perturbation technique. 
The stability of the triangular points is studied in stellar binary systems Kepler-34, Kepler-35, Kepler-413 and Kepler-16. To investigate the stability of the triangular points, we have obtained the expressions for critical mass which depends on the radiation of both primaries.
Fourier-series method is applied to obtain periodic orbits of the infinitesimal mass around triangular points in binary stellar systems.
We have obtained Fourier expansions of the periodic orbits around triangular points upto third order terms. A comparison is made between periodic orbits obtained by Fourier-series method and with Runge-Kutta integration of fourth order.
\end{abstract}

\section{Introduction}\label{sec:3sec1}
In a binary stellar system, each star is exerting a gravitational force on any object with mass in its vicinity.
The study of binary stellar systems is very important because of more than 60 percent of the stars in the solar neighborhood build such systems which are interesting systems from the dynamical point of view. For such system the restricted three body problem is an appropriate model.  
Restricted three body problem is a special case of the three body problem \citep{barrow1999poincar,roy2005orbital,marchal2012three,Mia2016MNRAS.457.1089M} which is very important in many scientific fields such as celestial mechanics, galactic dynamics, 
molecular theory etc. and it is the most important problem in exoplanetary systems.
In the circular restricted three body problem two primaries move in circular orbits around their common center of mass and the third body i.e., small body moves in the same plane and it is influenced by the gravitational forces from the primaries  but it does not influence 
the motion of the two primaries. So many authors have used this problem to find the stability of planet, asteroids or small bodies like meteoroids. 

Many researchers worked on the study of the existence and stability of libration points in the framework of restricted three body problem considering additional perturbations from radiation pressure \citep{alvarez2014stability,eapen2014study}, the stokes drag \citep{Jain2015Ap&SS.358...51J}, the solar wind drag \citep{pal2015geometry}, oblate and triaxial bodies \citep{singh2014effects}, the variable masses \citep{Abouelmagd2015Ap&SS.357...58A}, the potential from a disk \citep{Kishor2013MNRAS.436.1741K}, and the Yarkovsky effect \citep{ershkov2012yarkovsky}. 

There are several authors who have studied the significance of radiation from stars in binary systems \citep{Zhou1988Ap&SS.141..257Z,Papadakis2006Ap&SS.305...57P,roman2007photogravitational,
das2008effect}. Also, some authors studied the effect of radiation  from Sun on the motion of small particle in our solar system \citep{Robertson1937MNRAS..97..423R,Kushvah2008Ap&SS.315..231K,Abouelmagd2013Ap&SS.344..321A,
Kumari2013Ap&SS.344..347K,tiwary2015computation}. 
\cite{Todoran1993Ap&SS.201..281T} studied the effects of the radiation pressure in the restricted three-body problem and the existence of the `out-of-plane' equilibrium points. They found that within the framework of the stellar stability, the five Lagrangian points are the only
equilibrium points, at least as far as the force of the radiation pressure is taken into account.

The linear stability of the triangular equilibrium points in the photogravitational elliptic restricted problem have examined by \cite{Markellos1992Ap&SS.194..207M}. They determined the stability regions in the space of the parameters of mass, eccentricity, and radiation pressure. They also found that radiation pressure of the larger body for solar system case exerts only a small quantitative influence on the stability regions.

\cite{schwarz2012stability} studied the stability of the Lagrangian point L4 in the spatial restricted three-body problem and the possibility of inclined Trojan-like objects in exoplanetary systems (single and binary star systems). They investigated stability computing stability maps by numerical methods. In the case of circular motion of the primary bodies, they have shown that there are stable orbits up to an inclination $i = 61$ degree of the test particle.

\cite{abouelmagd2013stability} studied the existence of triangular points and their linear stability
when the primaries are oblate spheroid and sources of radiation in the restricted three-body problem. They observed that the
locations of the triangular points are affected by the oblateness of the primaries and solar
radiation pressure. They also shown that these points are stable for $0\leq\mu\leq\mu_c$ and unstable for $\mu_c\leq\mu\leq\frac{1}{2}$, where $\mu_c$ is the critical value of mass. 
Recently, \cite{Bosanac2015CeMDA.122...27B} investigated periodic motions near a large mass ratio binary system within the context
of the circular restricted three-body problem. They used stability analysis to explore the effect of the mass ratio on the structure of families of periodic orbits.

But most of their works includes the use of $q_i=1-\beta_i, i=1,2$, where $\beta_{i}$ is the ratio of the radiation pressure force of binary components to gravitational force  of  binary components. The luminosity of some star of binary's is unknown so in this present work using relation between luminosity and mass of  a star, we have calculated the values of luminosity for different binary systems. Also using the realistic relation among $\beta$, luminosity and mass of the binary components, we have found the values of $q$ for different binary system. After that we study the motion of a small particle around a binary star system in the presence of radiation from both the stars i.e., primaries. 

Moreover, we have found periodic orbits around triangular points with the aid of Fourier series. Periodic orbits with the help of Fourier series of the classical RTBP have found by \cite{pedersen1935fourier}. Whereas, we determine periodic orbits of the RTBP with radiating primaries in binary systems. Also for the purpose of matching the results we have determined periodic orbits in Earth-Moon system and the result found agree with that of \cite{pedersen1935fourier}. 
 
The purpose of this present study is to describe and examine the motion and to find the orbits of the infinitesimal body in stellar binary system using the model of circular restricted three body problem taking into account of the photo gravitational effects of both the stars along with the gravitational forces. We  analyze the stability of triangular points in the binary stellar system (Kepler-34, Kepler-35, Kepler-413 and Kepler-16) by the method described in \cite{moulton2012introduction,Szebehely1967torp.book.....S,Murray2000ssd..book.....M}. 
We obtained periodic orbits around triangular points with the help of Fourier series.

The present paper  is organized as follows. In Section \ref{sec:3sec2}, we introduce the equations of motion. Numerical values of physical parameters are discussed in Section \ref{sec:3sec3}. In Section \ref{sec:3sec4}, we found location of Lagrangian points of binary stellar systems. Stability of triangular points are discussed in   Section \ref{sec:3sec5}. Fourier series in the vicinity of triangular points are discussed in Section \ref{sec:3sec7}. Finally, Section \ref{sec:3sec8} is devoted to conclusions.

\section{The equations of motion} \label{sec:3sec2}
We assume that $M^{\ast}_1$  and  $M^{\ast}_2$ be the masses of primary and secondary star having spherical symmetry move about their centre of mass in circular orbits and  a third mass also known as infinitesimal body $m$ attracted by the previous two but not influencing their motion. We consider the motion of a test particle under the influence of the gravitational force of the star-star system, and $q_1$ and $q_2$, the factor characterizing the radiation effect of the primary and secondary stars respectively. We have chosen the unit of mass such that $M^{\ast}_1+M^{\ast}_2=1$. Let unit of distance be so chosen such that the constant distance between two finite bodies is unity. We consider the mass parameter $\mu =\frac{M^{\ast}_2}{M^{\ast}_1+M^{\ast}_2}$, let unit of time be so chosen that $G$ shall be equal to unity and mean motion is
$n=1$ with $M^{\ast}_2=\mu$ and $M^{\ast}_1=1-\mu$. Let us suppose that $Oxyz$ be the rotating co-ordinate system having origin at the centre of mass $O$ of the primaries which is fixed with respect to the inertial system. Let the position of infinitesimal body be $P(x,y)$ and the positions of bigger and smaller bodies are $P_{1}(-\mu,0)$ and $P_{2}(1-\mu,0)$ respectively relative  to the rotating system. Then the equations of motion of the test particle in the dimensionless rotating coordinate system are written as \citep{Zagouras1991CeMDA..51..331Z}
\begin{eqnarray}
\ddot x-2\dot y=\Omega_x = x-\frac{q_1(1-\mu)(x+\mu)}{r_1^3}-\frac{q_2\mu(x+\mu- 1)}{r_2^3},\label{eq:3eq1}\\
\ddot y+2\dot x=\Omega_y = y-\frac{q_1(1-\mu)y}{r_1^3}-\frac{q_2\mu y}{r_2^3},~~~~~~~~~~~~~~~~~~~~~ \label{eq:3eq2}
\end{eqnarray} 
where
\begin{eqnarray}
&&\Omega=\frac{1}{2}(x^2+y^2)+\frac{q_1(1-\mu)}{r_1}+\frac{q_2\mu}{r_2},\\&&
r_1^2=(x+\mu)^2+y^2, \qquad r_2^2=(x+\mu-1)^2+y^2.
\end{eqnarray}
Also the Jacobi integral is given by 
\begin{eqnarray}
&&\dot x^2+\dot y^2=2\Omega -C,\label{eq:3eq9}
\end{eqnarray}
where $C$ is the Jacobi constant.

\section{Numerical values of physical parameters}\label{sec:3sec3}
The mass reduction factor $q$ is defined as $q=1-\beta$, where the parameter $\beta=\frac{F_p}{F_g}$, $F_p$ and $F_g$ be the radiation and gravitational attraction forces respectively. It depends on the physical properties of the star and the test particle which is given by  \citep{Ragos1993Ap&SS.209..267R}
\begin{eqnarray}
&&\beta=\frac{3L}{16\pi c GM\rho s},
\end{eqnarray}
where $M $ and $L$ are the mass and luminosity of a star,  $s$   and  $\rho$ are the radius and density of the particle and $c$ and $G$ are the speed of light and Gravitational constant respectively. \cite{brownlee1976physical} have shown that micron-sized interplanetary dust particles have densities between 1 g/cm$^3$ and 4 g/cm$^3$. The average density of cosmic dust is typically around 2 g/cm$^3$ with most values between 0.6 and 5.5 g/cm$^3$ \citep{kohout2014density}. Moreover, \cite{singh2013out} considered $s=2\times10^{-2}$ cm and $\rho=1.4$g/cm$^3$ for some dust grain particles in the binary systems. Likewise in this paper we have considered $s=7\times 10^{-3}$cm and $\rho=1.5$ g/cm$^3$. Also for stars, luminosity is related to mass as \citep{duric2004advanced,salaris2005evolution}
\begin{eqnarray}
 &&\frac{L}{L_\odot}\approx (\frac{M}{M_\odot})^{3.9},
 \end{eqnarray}
 where, $L_\odot$ and $M_\odot$ are the luminosity  and  mass of the Sun. We compute the values of luminosity $L$ and mass reduction factor $q$ using the parameters values in C.G.S units as
$L_\odot=3.846\times 10^{33}$ erg/s, $c=3\times 10^{10}$ cm/s, $G=6.67384\times 10^{-8}$ cm$^{3}$g$^{-1}$s$^{-2}$ and $M_{\odot}=1.99\times 10^{33}$g  and computed values are given in Table \ref{tab:1}.

\begin{table*}[h]
 \centering
 \begin{minipage}{120mm}
  \caption{Physical parameters of four binary candidates. \label{tab:1}}
  \begin{tabular}{llccccccccc}
  \hline
  System  &$M^{\ast}_1(M_\odot)$ &$M^{\ast}_2(M_\odot)$ &$\mu$ &$L^{\ast}_1(L_\odot)$&$L^{\ast}_2(L_\odot)$&$q_1$ & $q_2$ &\tabularnewline
  \hline
Kepler 34 & 1.0479 &1.0208  & 0.49345   &1.20018 &1.083620 &0.993716 & 0.994176 &\tabularnewline
Kepler 35 & 0.8877 &0.8094  & 0.476931 &0.628403 &0.438366 &0.996116 & 0.997028 &\tabularnewline        
Kepler 413 &0.82 &0.5423  & 0.398077&0.461184 &0.091946 &0.996914 & 0.999070&\tabularnewline
Kepler 16 &0.6897 &0.20255& 0.22701&0.234842 &0.00197458 &0.998132 & 0.999947&\tabularnewline
\hline
\end{tabular}
\end{minipage}
\end{table*}
 
 \section{Location of the equilibrium points in binary star systems}\label{sec:3sec4}
There are five equilibrium points also known as Lagrangian points for the restricted three body problem. 
At the equilibrium  points $\dot x=\dot y= \ddot x=\ddot y=0$. So, the co-ordinates of equilibrium  points of the problem are obtained by equating R.H.S of Eqs. (\ref{eq:3eq1}) and (\ref{eq:3eq2})  to zero i.e.,
\begin{eqnarray}
&&x-\frac{q_1(1-\mu)(x+\mu)}{r_1^3}-\frac{q_2\mu(x+\mu- 1)}{r_2^3}=0,\label{eq:3eq3}\\
&&y-\frac{q_1(1-\mu)y}{r_1^3}-\frac{q_2\mu y}{r_2^3}=0. \label{eq:3eq4}
\end{eqnarray} 
In the classical RTBP there are five equilibrium points namely $L_1, L_2, L_3 , L_4 $ and $L_5$ and depend on the masses of the respective primaries. But the location of equilibrium points in binary star system depends on the masses as well as the luminosities of the stars. 
\subsection{Location of the collinear points}
At the collinear points, $y=0, \quad r_1=|x+\mu|, \quad$ and \quad $r_2=|x-(1-\mu)|$. There are three such points $L_1, L_2$ and $L_3$, where $L_1$ lies between $M^{\ast}_1$ and $M^{\ast}_2$, $L_2$ lies to the right side of mass $M^{\ast}_2$, $L_3$ lies to the left side of mass $M^{\ast}_1$.
At $L_1$ point, $-\mu<x<1-\mu$. So, in this case, we have $r_2=1-\mu-x=\rho$(say), $r_1=1-\rho$ and $x=1-\mu-\rho$.
On substituting these values in Eq. (\ref{eq:3eq3}), we get
\begin{eqnarray}
&&\rho^5-(3-\mu)\rho^4+(3-2\mu)\rho^3-(1-q_1-\mu+q_1\mu+q_2\mu)\rho^2\nonumber\\
&&~~~~~~~~~~~~~~~~~~~~~~~~~~~~~~~~~~~~~~~~~~~~~~~~~~+2q_2\mu\rho-q_2\mu=0.\label{eq:3eqc1}
\end{eqnarray}
For the classical restricted three body problem, when $q_1=q_2=1$, we assume that $\gamma$ be the value of $\rho$ where, $\gamma$
is given by \citep{fitzpatrick2012introduction}
\begin{eqnarray}
\gamma=\alpha-\frac{\alpha^2}{3}-\frac{\alpha^3}{9}-\frac{23}{81}\alpha^4+\mathcal{O}(\alpha^5), \quad \alpha=\left(\frac{\mu}{3(1-\mu)}\right)^\frac{1}{3}.
\end{eqnarray}
Suppose the value of $\rho$  slightly changes in the presence of $q_1$ and $q_2$ and new value of $\rho$ is defined as $\rho=\gamma+\delta_1$, where $\delta_1\ll 1$. We determine the value of $\delta_1$ by substituting the value of $\rho$ the Eq. (\ref{eq:3eqc1}) and solving them (after neglecting the higher order terms of $\delta_1$ as $\delta_1\ll 1$), we obtain
\begin{eqnarray}
\delta_1=\frac{A_1+A_2 q_1+A_3 q_2}{A_4+A_5 q_1+A_6 q_2},
\end{eqnarray}
where, 
\begin{eqnarray}
&&A_1=-\gamma^5+(3-\mu)\gamma^4-(3-2\mu)\gamma^3+(1-\mu)\gamma^2, \ A_2=-\gamma^2(1-\mu),\nonumber\\
&&A_3=\mu(1-\gamma)^2, \  A_4=5\gamma^4-4(3-\mu)\gamma^3+3(3-2\mu)\gamma^2-2(1-\mu)\gamma,\nonumber\\
&&A_5=2\gamma(1-\mu), \ A_6=2\mu(1-\gamma).
\end{eqnarray}
At $L_2$ points, $x>1-\mu$. In this case, we have $r_2=x+\mu-1=\rho \ (\mbox{say}),\ r_1=1+\rho \ \mbox{and}\ x=1-\mu+\rho$. Substituting these values in Eq. (\ref{eq:3eq3}) and simplifying, we have
\begin{eqnarray}
&&\rho^5+(3-\mu)\rho^4+(3-2\mu)\rho^3+(1-q_1-\mu+q_1 \mu-q_2\mu)\rho^2\nonumber\\
&&~~~~~~~~~~~~~~~~~~~~~~~~~~~~~~~~~~~~~~~~~~~~~~~~~~~-2q_2 \mu \rho-q_2\mu=0.\label{eq:3eqc2}
\end{eqnarray}
Suppose $\gamma$ be the value of $\rho$ for classical case and it is given by \citep{fitzpatrick2012introduction}
\begin{eqnarray}
\gamma=\alpha+\frac{\alpha^2}{3}-\frac{\alpha^3}{9}-\frac{31}{81}\alpha^4+\mathcal{O}(\alpha^5), \quad \alpha=\left(\frac{\mu}{3(1-\mu)}\right)^\frac{1}{3}.
\end{eqnarray}
For the presence of $q_1$ and $q_2$, let the value of $\rho$ will be slightly changed and the new value is defined as 
\begin{eqnarray}
\rho=\gamma+\delta_2, \qquad \mbox{where} \ \delta_2\ll1.
\end{eqnarray}
 Now, substituting this value of $\rho$ in Eq. (\ref{eq:3eqc2}) and neglecting the higher order terms of $\delta_2$ as $\delta_2\ll1$, we determine
\begin{eqnarray}
\delta_2=\frac{B_1+B_2 q_1+B_3 q_2}{B_4+B_5 q_1+B_6 q_2},
\end{eqnarray}
where, 
\begin{eqnarray}
&&B_1=-\gamma^5-(3-\mu)\gamma^4-(3-2\mu)\gamma^3-(1-\mu)\gamma^2,\ B_2=\gamma^2(1-\mu),\nonumber\\
&&B_3=\mu(1+\gamma)^2, \ B_4=5\gamma^4+(12-4\mu)\gamma^3+(9-6\mu)\gamma^2+2(1-\mu)\gamma,\nonumber\\
&&B_5=-2\gamma(1-\mu), \ B_6=-2\mu(1+\gamma).
\end{eqnarray}
Similarly, at $L_3$ points, $-\infty<x<-\mu$. In this case, we have $r_1=-(x+\mu)=\rho$ (say), $r_2=1+\rho,$ and $x=-\rho-\mu.$ On substituting these values in Eq. (\ref{eq:3eq3}) and after simplifying, we have
\begin{eqnarray}
&&\rho^5+(2+\mu)\rho^4+(1+2\mu)\rho^3+(-q_1+\mu+q_1\mu-q_2\mu)\rho^2+(-2q_1+2q_1\mu)\rho\nonumber\\
&&~~~~~~~~~~~~~~~~~~~~~~~~~~~~~~~~~~~~~~~~~~~~~~~~~~~~~~~~~~~~~~~-q_1+q_1\mu=0.\label{eq:3eqc3}
\end{eqnarray}
In this case, also assuming that, $\beta$ be the classical value of $\rho$ and it is given as \citep{fitzpatrick2012introduction}
\begin{eqnarray}
\beta=\frac{7}{12}\frac{\mu}{(1-\mu)}-\frac{7}{12}\left(\frac{\mu}{1-\mu}\right)^2+\frac{13223}{20736}\left(\frac{\mu}{1-\mu}\right)^3+\mathcal{O}(\alpha^4).
\end{eqnarray}
For the presence of $q_1$ and $q_2$, let the value of $\rho$ will be slightly changed and the new value is defined as 
\begin{eqnarray}
\rho=\gamma+\delta_3, \qquad \mbox{where} \ \delta_3\ll1.
\end{eqnarray}
 On substituting this value of $\rho$ in Eq. (\ref{eq:3eqc3}) and neglecting the higher order terms of $\delta_3$ as $\delta_3\ll1$, we determine
\begin{eqnarray}
\delta_3=\frac{C_1+C_2 q_1+C_3 q_2}{C_4+C_5 q_1+C_6 q_2},
\end{eqnarray}
where,
\begin{eqnarray}
&&\ C_1=-\beta^5-(2+\mu)\beta^4-(1+2\mu)\beta^3-\beta^2\mu, \ C_2=(1+\beta)^2(1-\mu),\nonumber\\
&&C_3=\beta^2\mu, \ C_4=5\beta^4+(8+4\mu)\beta^3+(3+6\mu)\beta^2+2\beta\mu,\nonumber\\
&&C_5=-2(1+\beta)(1-\mu), \ C_6=-2\beta\mu.
\end{eqnarray} 
Using the above expressions we can obtain the collinear points for any binary star systems. For the purpose of validation of the semi-analytical result we have solved numerically  Eq. \eqref{eq:3eq3} and obtained collinear points $L_1$ and $L_2$ for four binary star systems. From Table \ref{tab:2a}, it is clear that semi-analytical and numerical results shows an excellent level of agreement.  
\begin{table*}[h]
 \centering
 \begin{minipage}{120mm}
  \caption{Comparison of collinear points $L_1$ and $L_2$ obtain by semi-analytical and numerical method. \label{tab:2a}}
  \begin{tabular}{llccccccccc}
  \hline
  System &$L_1$(anlytical)& $L_1$(numerical)&$L_2$(anlytical)& $L_2$(numerical)&\tabularnewline
  \hline
Kepler 34 &0.0088234&0.0091894&1.20114&1.19897&\tabularnewline
Kepler 35 &0.0319996&0.0325072&1.20731&1.20461&\tabularnewline        
Kepler 413&0.1435990&0.1442520&1.23213&1.22975& \tabularnewline
Kepler 16 &0.3955430&0.3956840&1.26888&1.26737&\tabularnewline
\hline
\end{tabular}
\end{minipage}
\end{table*}
\subsection{Location of the triangular points}
\begin{figure*}[h]
 \begin{center}
 \includegraphics[width=0.56\textwidth]{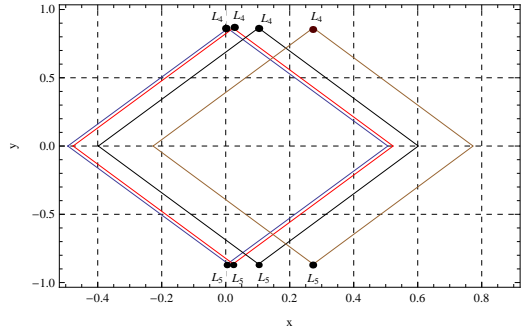}
\caption{Location of triangular points from left to right: Kepler-34, Kepler-35, Kepler-413, Kepler-16. \label{fig:4f1}}
 \end{center}
 \end{figure*}

The triangular points $L_4$ and $L_5$ have the coordinates \citep{Schuerman1980ApJ...238..337S}
\begin{eqnarray}
&&x=\frac{1}{2}(1+q_2^{\frac{2}{3}}-q_1^{\frac{2}{3}})-\mu\label{eq:3eq7},\\
&&y=\pm\frac{1}{2}[4q_1^{\frac{2}{3}}q_2^{\frac{2}{3}}-(q_1^{\frac{2}{3}}+q_2^{\frac{2}{3}}-1)^2]^{\frac{1}{2}}.\label{eq:3eq8}
\end{eqnarray}
Using Eqs. \eqref{eq:3eq7} and \eqref{eq:3eq8}, we obtain 
the position of triangular equilibrium points  of the binary star systems. The computed  numerical values of the triangular points are  given in Table \ref{tab:2b} and a comparison of triangular points of different systems are shown in Fig. \ref{fig:4f1}.
\begin{table*}[h]
 \centering
 \begin{minipage}{120mm}
  \caption{Triangular points $L_4(x,y)$ and $L_5(x,-y)$. \label{tab:2b}}
  \begin{tabular}{llccccccccc}
  \hline
  System &$q_1$ & $q_2$&$x$ &$y$&\tabularnewline
  \hline
Kepler 34 &0.993716&0.994176& 0.00670364 &0.860897&\tabularnewline
Kepler 35 &0.996116&0.997028& 0.0233733 &0.863277&\tabularnewline        
Kepler 413&0.996914&0.999070&0.102642 &0.864805&\tabularnewline
Kepler 16 &0.998132&0.999947&0.273595 &0.86563&\tabularnewline
\hline
\end{tabular}
\end{minipage}
\end{table*}

\section{Stability of the triangular points}\label{sec:3sec5}
We are now interested to know what would happen if the infinitesimal particle is displaced a little from one of the triangular points. If the infinitesimal particle is slightly displaced from one of the equilibrium points and given a small velocity to that particle then either the motion of the particle is a rapid departure from the vicinity of the point, we call such a position of equilibrium as an unstable 
one or the particle merely oscillates about the point, this position is known as stable position.
In order to study the possible  motion of the infinitesimal particle near any Lagrangian point $(x_{0}, \ y_{0})$,  the particle be displaced to the point
$(x_{0}+\xi, \ y_{0}+\eta)$.
 We define 
\begin{eqnarray}\label{eq:10}
&&x=x_{0}+\xi, \ y=y_{0}+\eta,
\end{eqnarray}
 where the displacements 
$\xi=Ae^{\lambda t}, \ \eta=Be^{\lambda t}$ are very small, $A, \ B$ and $\lambda$ are parameters to be determined. Putting these 
co-ordinates into Eqs. (\ref{eq:3eq1}) and (\ref{eq:3eq2}) and using the method discussed by \cite{Murray2000ssd..book.....M}, we 
obtain
\begin{eqnarray}
 &&\ddot{\xi}-2\dot{\eta}=\xi \Omega^0_{xx}+\eta \Omega^0_{xy} ,\label{eq:51}\\&&
 \ddot{\eta}+2\dot{\xi}=\xi \Omega^0_{xy}+\eta \Omega^0_{yy},\label{eq:52}
\end{eqnarray} 
 where superfix $0$ denotes corresponding value at equilibrium point. Putting $\xi=Ae^{\lambda t}, \eta=Be^{\lambda t}$ and 
after simplifying, we obtain the characteristic equation corresponding to Eqs. (\ref{eq:51}) and (\ref{eq:52}) as
\begin{eqnarray}
&&\lambda^4+(4-\Omega^0_{xx}-\Omega^0_{yy})\lambda^2+\Omega^0_{xx}\Omega^0_{yy}
-(\Omega^0_{xy})^2=0.\label{eq:3eq14}
\end{eqnarray}
Let us suppose that $q_i=1-\beta_i (i=1,2)$. Again as $0<q_i\leq1$, we have $\beta_1, \beta_2\ll1$.
Now, we have
\begin{eqnarray}
&&\Omega^0_{xx}=\frac{3}{4}+h_1+\mu k_1, \nonumber\\
&&\Omega^0_{yy}=\frac{9}{4}+h_2+\mu k_2,\nonumber\\
&&\Omega^0_{xy}=\frac{3}{4}\sqrt{3}(1-2\mu)+h_3+\mu k_3,\nonumber\\
&&\mbox{with}\ h_1=\frac{3}{2}\beta_1-\beta_2, \ k_1=\frac{5}{2}(\beta_2-\beta_1),\ h_2=\frac{1}{2}\beta_1-\beta_2,\nonumber\\
&&k_2=\frac{3}{2}(\beta_2-\beta_1),\ h_3=\sqrt{3}(\frac{5}{6}\beta_1-\frac{2}{3}\beta_2), \ k_3=-\frac{\sqrt{3}}{6}(\beta_1+\beta_2).\nonumber
\end{eqnarray}
As $\beta_1, \beta_2\ll1$, each of $|h_i|$ and $|k_i|~~ (i=1,2,3)$ is very small.
It is noticed that we neglect the higher order terms containing $\beta_1$ and $\beta_2$ as $\beta_1, \beta_2\ll1$. On substituting these values of $\Omega^0_{xx}$, $\Omega^0_{yy}$ and $\Omega^0_{xy}$ in Eq. (\ref{eq:3eq14}), the characteristic equation becomes
\begin{eqnarray}
\lambda^4+B\lambda^2+C=0,\label{eq:3rev1}
\end{eqnarray} 
where,
\begin{eqnarray}
&&B=1-h_1-h_2-(k_1+k_2)\mu,\nonumber\\
&&C=\frac{9}{4}h_1+\frac{3}{4}h_2-\frac{3\sqrt{3}}{2}h_3+(\frac{27}{4}+3\sqrt{3}h_3+\frac{9}{4}k_1+\frac{3}{4}k_2-\frac{3\sqrt{3}}{2}k_3)\mu\nonumber\\
&&~~~~~+(-\frac{27}{4}+3\sqrt{3}k_3)\mu^2.\nonumber
\end{eqnarray}
The roots of Eq. (\ref{eq:3rev1}) are 
\begin{eqnarray}
\lambda^2=\frac{-B\pm \sqrt{B^2-4C}}{2},\label{eq:3rev2}
\end{eqnarray}
where, the discriminant of Eq. (\ref{eq:3rev2}) is given by 
\begin{eqnarray}
&&B^2-4C=1-11h_1-5h_2+6\sqrt{3}h_3+(-27-12\sqrt{3}h_3-11k_1-5k_2+6\sqrt{3}k_3)\mu\nonumber\\
&&~~~~~~~~~~~~~~~~+(27-12\sqrt{3}k_3)\mu^2.
\end{eqnarray}
The critical value of mass parameter $\mu_{c}$ is a root of the equation $B^2-4C=0$, solving this equation, we obtain 
\begin{eqnarray}
&&\mu_{c}=\frac{1}{2}-\frac{1}{6}\sqrt{\frac{23}{3}}-\frac{11}{3\sqrt{69}}h_1-\frac{5}{3\sqrt{69}}h_2+\frac{2}{3\sqrt{3}}h_3\nonumber\\
&&~~~~~~~~~+\left(\frac{11}{54}-\frac{11}{6\sqrt{69}}\right)k_1+\left(\frac{5}{54}-\frac{5}{6\sqrt{69}}\right)k_2+\left(\frac{1}{3\sqrt{3}}-\frac{\sqrt{23}}{27}\right)k_3.
\end{eqnarray}
Now for classical restricted three body problem, $q_1=q_2=1$ and hence $\beta_1=\beta_2=0$. Consequently, $h_i=k_i=0 ~~(i=1,2,3)$ and in this case, the critical value of the mass ratio is  $\mu_c=\frac{1}{2}-\frac{1}{6}\sqrt{\frac{23}{3}}=0.0385209$ which is same as that of classical case \citep{Szebehely1967torp.book.....S,celletti2010stability}. There can be three cases on the basis of the critical value of mass parameter which are  
\begin{itemize}
\item when $\mu_c<\mu\leq\frac{1}{2}$ or $B^2-4C<0$, the real part of two of the characteristic roots are positive and equal and hence in this case triangular points are unstable.
\item  when $0<\mu<\mu_c$ or $B^2-4C>0$, all the four roots of characteristic equation are pure imaginary and different and consequently triangular points are stable.
\item when $\mu=\mu_c$ or $B^2-4C=0$,  there are double roots of equal magnitude which means secular terms present in the solutions of the variational equation  and hence triangular points are unstable.
\end{itemize}
In our cases the value of $\mu_c$ for four binary systems are given in Table \ref{tab:3tab3}.
We have seen from Table \ref{tab:3tab3} that for all the systems, critical value $\mu_c $ of mass parameter satisfies the relation $\mu_c<\mu<\frac{1}{2}$, and the discriminant of Eq. \eqref{eq:3rev2}  $B^2-4C$ is negative.
Hence, the real part of two of the characteristic roots are positive and equal which are given in Table \ref{tab:3tab3}. Consequently, the triangular equilibrium points are unstable.
\begin{table*}[h]
 \centering
 \begin{minipage}{120mm}
  \caption{Critical values of mass parameter and eigen values at $L_{4,5}$. \label{tab:3tab3}}
  \begin{tabular}{llccccccccc}
  \hline
  System  & $q_1$ & $q_2$ &$\mu_c$&$\lambda_{1,2}$&$\lambda_{3,4}$&\tabularnewline
  \hline
Kepler 34 &0.993716 & 0.994176 &0.0383448&0.632724$\pm$ 0.948859i&$-0.632724\pm$ 0.948859i&\\
Kepler 35  &0.996116 & 0.997028 &0.0383246&0.631936$\pm$ 0.948315i&$-0.631936\pm$ 0.948315i&\\
Kepler 413  &0.996914 & 0.99907&0.0381657&0.621598$\pm$ 0.941247i&$-0.621598\pm$ 0.941247i&\\
Kepler 16  &0.998132 & 0.999947&0.0382349&0.542934$\pm$ 0.890947i&$-0.542934\pm$ 0.890947i&\\
\hline
\end{tabular}
\end{minipage}
\end{table*}

\section{Fourier series for the periodic orbits in the vicinity of triangular points}\label{sec:3sec7}
In mathematics, a Fourier series decomposes any periodic function into the sum of a possibly infinite set of oscillating functions, namely sines and cosines. Fourier series were introduced by Joseph Fourier for the purpose of solving the heat equation in a metal plate. Though the original motivation was to solve the heat equation, later it became clear that the same techniques could be applied to a broad range of mathematical and physical problems.
A periodic orbit is a special type of solution for a dynamical system, which repeats itself in time. In the field of astronomy, astrophysics, and  space science etc., periodic orbits play much important role and from their studies one may know about orbital resonance, spin orbit etc. In this section we determine periodic orbits around triangular points using Fourier series.
To discuss the motion of the infinitesimal mass in the vicinity of the triangular points $L_4$, we translate the origin to $L_4$ point with new co-ordinate system whose axes parallel to old co-ordinate system. Then we have
\begin{eqnarray}
\begin{cases}
x=\xi+x_0=\xi+\frac{1}{2}(1+q_2^{\frac{2}{3}}-q_1^{\frac{2}{3}})-\mu,\\
y=\eta+y_0=\eta+\frac{1}{2}[4q_1^{\frac{2}{3}}q_2^{\frac{2}{3}}-(q_1^{\frac{2}{3}}+q_2^{\frac{2}{3}}-1)^2]^{\frac{1}{2}}.
\end{cases}
\end{eqnarray} 
 The equations of motion in new co-ordinate system are
 \begin{eqnarray}
 &&\ddot \xi-2\dot \eta-\xi-\frac{h}{2}+\mu=\frac{\partial \Omega_f}{\partial \xi},\label{eq:3eq61}\\
 &&\ddot \eta-2\dot \xi-\eta-y_0=\frac{\partial \Omega_f}{\partial \eta},\label{eq:3eq62}
 \end{eqnarray}
 where
 \begin{eqnarray}
&&\Omega_f=\frac{q_1(1-\mu)}{r_1}+\frac{q_2\mu}{r_2},\label{eq:3eq63}\\
&&h=1+q_2^{\frac{2}{3}}-q_1^{\frac{2}{3}},\\ 
&&r_1^2=(\xi+\frac{h}{2})^2+(\eta+y_0)^2,\\
&&r_2^2=(\xi+\frac{h}{2}-1)^2+(\eta+y_0)^2,
 \end{eqnarray}
We expand Eq.(\ref{eq:3eq63}) upto fourth order terms as follows:
\begin{eqnarray}
&&\Omega_f=A_{00}+A_{10}\xi+A_{01}\eta+A_{20}\xi^2+A_{11}\xi\eta+A_{02}\eta^2
+A_{30}\xi^3+A_{21}\xi^2\eta\nonumber\\
&&~~~~~~+A_{12}\xi\eta^2+A_{03}\eta^3+A_{40}\xi^4+A_{31}\xi^3\eta+A_{22}\xi^2\eta^2+A_{13}\xi\eta^3+A_{04}\eta^4,\label{eq:3eq67}
\end{eqnarray}
where the coefficients of the right hand side are given in Appendix A.
From Eq.(\ref{eq:3eq67}) we can find the expansions of $\frac{\partial \Omega_f}{\partial \xi}$ and 
$\frac{\partial \Omega_f}{\partial \eta}$ and then inserted in Eqs.(\ref{eq:3eq61}) and (\ref{eq:3eq62}), we obtained the governing equations of motion of the infinitesimal mass in the vicinity of triangular 
point $L_4$ upto third order terms as follows:
\begin{eqnarray}
&&\ddot{\xi}-2\dot{\eta}-\frac{h}{2}+\mu-A_{10}-(1+2A_{20})\xi-A_{11}\eta-3A_{30}\xi^2
-2A_{21}\xi\eta-A_{12}\eta^2\nonumber\\
&&~~~~~~~~~~~~~~~~~~~~~~~~-4A_{40}\xi^3-3A_{31}\xi^2\eta-2A_{22}\xi\eta^2-A_{13}\eta^3=0,\label{eq:3eq68}\\
&&\ddot{\eta}+2\dot{\xi}-y_0-A_{01}-A_{11}\xi-(1+2A_{02})\eta-A_{21}\xi^2
-2A_{12}\xi\eta\nonumber\\
&&~~~~~~~~~~~-3A_{03}\eta^2-A_{31}\xi^3-2A_{22}\xi^2\eta-3A_{13}\xi\eta^2-4A_{04}\eta^3=0,\label{eq:3eq69}
\end{eqnarray}
and the Jacobi integral corresponding to Eqs.(\ref{eq:3eq68}) and (\ref{eq:3eq69}) is 
\begin{eqnarray}
&&\dot{\xi}^2+\dot{\eta}^2=(h-2\mu+2A_{10})\xi+(2y_0+2A_{01})\eta
+(1+2A_{02})\eta^2+(1+2A_{20})\xi^2\nonumber\\
&&~~~~+2A_{00}+2A_{03}\eta^3+2A_{04}\eta^4+2A_{11}\xi\eta+2A_{12}\xi\eta^2
+2A_{13}\xi\eta^3+2A_{21}\xi^2\eta\nonumber\\
&&~~~+2A_{22}\xi^2\eta^2+2A_{30}\xi^3+2A_{31}\xi^3\eta+2A_{40}\xi^4-C.
\end{eqnarray}
To find the periodic orbits around triangular point $L_4$, we have to determine the Fourier coefficients
$a$ and $b$ in the Fourier expansions of $\xi$ and $\eta$:
\begin{eqnarray}
&&\xi=a_0+\sum_{j=1}^3{a_j\cos j\omega t}+\sum_{j=1}^{3}{a_{-j}\sin j\omega t},\label{eq:3eq71}\\
&&\eta=b_0+\sum_{j=1}^3{b_j\cos j\omega t}+\sum_{j=1}^{3}{b_{-j}\sin j\omega t},\label{eq:3eq72}
\end{eqnarray} 
where the coefficients  $a_1, a_{-1}, b_1,$ and $b_{-1}$ are of the first order, while the remaining  coefficients are of higher order. The coefficients with subscripts $0, 2$ and $-2$ are of the second order while the coefficients with subscripts $3$ and $-3$ are of third order. 
\subsection{First order Fourier coefficients}
Keeping the first order terms only in Eqs.(\ref{eq:3eq68}), (\ref{eq:3eq69}), (\ref{eq:3eq71}) and
(\ref{eq:3eq72}) and with the help of these equations, we have four equations for determining the first order terms as follows:
\begin{eqnarray}
\begin{cases}
(1+2A_{20}+\omega^2)a_1+A_{11}b_1+2\omega b_{-1}=0,\\
A_{11}a_1+(1+2A_{02}+\omega^2)b_1-2\omega a_{-1}=0,\\\label{eq:3eq73}
-2\omega b_1+(1+2A_{20}+\omega^2)a_{-1}+A_{11}b_{-1}=0,\\
2\omega a_1+A_{11}a_{-1}+(1+2A_{02}+\omega^2)b_{-1}=0.
\end{cases}
\end{eqnarray} 
We can determine $a_1$ and $b_1$ in terms of $a_1$ and $b_1$ from the first two equations of (\ref{eq:3eq73}) and then inserting these expressions of $a_{-1}$ and $b_{-1}$ in the last two equations of (\ref{eq:3eq73}), we obtain
\begin{eqnarray}
&&\left\{\omega^4+(-2+2A_{02}+2A_{20})\omega^2
+(1+2A_{20})(1+2A_{02})-A_{11}^2\right\}a_1=0,\label{eq:3eq74}\\
&&\left\{\omega^4+(-2+2A_{02}+2A_{20})\omega^2
+(1+2A_{20})(1+2A_{02})-A_{11}^2\right\}b_1=0,\label{eq:3eq75}
\end{eqnarray}  
From above two equations it is clear that if 
\begin{eqnarray}
&&\left\{\omega^4+(-2+2A_{02}+2A_{20})\omega^2
+(1+2A_{20})(1+2A_{02})-A_{11}^2\right\}< \mbox{or} > 0,
\end{eqnarray}
then $a_1=b_1=0$ and consequently $a_{-1}=b_{-1}=0$.
Therefore the mathematical conditions for periodic orbits is
\begin{eqnarray}
&&\left\{\omega^4+(-2+2A_{02}+2A_{20})\omega^2
+(1+2A_{20})(1+2A_{02})-A_{11}^2\right\}=0.
\end{eqnarray}
Now Eqs.(\ref{eq:3eq74}) and (\ref{eq:3eq75}) are satisfied for all values $a_1$ and $b_1$. We have chosen the values $a_1=\epsilon$ and $b_1=0$ \citep{pedersen1935fourier}.
Putting this values of $a_1$ and $b_1$ in (\ref{eq:3eq73}), we have
\begin{eqnarray}
\begin{cases}
a_{-1}=\frac{1}{2\omega}A_{11}\epsilon,\\
b_{-1}=-\frac{1}{2\omega}(1+2A_{20}+\omega^2)\epsilon.
\end{cases} 
\end{eqnarray}
Hence the periodic orbits around $L_4$ are
\begin{eqnarray}
&&\xi=\epsilon \cos\omega t+\frac{1}{2\omega}A_{11}\epsilon\sin\omega t,\\
&&\eta=-\frac{1}{2\omega}(1+2A_{20}+\omega^2)\epsilon\sin\omega t.
\end{eqnarray}
\subsection{Second order Fourier coefficients}
By applying a similar procedure as in the previous subsection, we obtain second order Fourier coefficients
of the periodic orbits around the triangular points $L_4$. For the following calculations we will use the coefficients scheme given in \cite{pedersen1933periodic}. Introducing the values of $a_1,\ b_1, \ a_{-1}, \ b_{-1}$ and $\xi, \ \eta, \ \xi^2, \ \eta^2, \ \xi\eta, \ \dot{\xi},\ \dot{\eta},\ \ddot{\xi},\ \ddot{\eta}$ and equating the coefficient of 1 from both sides of 
Eqs. (\ref{eq:3eq68}) and (\ref{eq:3eq69}) we get two equations with unknown $a_0$ and $b_0$ solving them, we obtain
\begin{eqnarray}
&&a_0=\frac{\epsilon^2(1+2A_{20}+\omega^2)}{8\omega^4\{2(-1+A_{02}+A_{20})+\omega^2\}}\left[3(1+2A_{02})(A_{12}+2A_{12}A_{20}-A_{11}A_{21}+A_{30}
\nonumber\right.\\&&\left.
+2A_{02}A_{30})
+\{{A_{12}(-3+6A_{02}+4A_{20})-A_{11}A_{21}+3A_{30}+6A_{02}A_{30}}\}\omega^2
\nonumber\right.\\&&\left.
+2A_{12}\omega^4-3A_{03}A_{11}(1+2A_{20}+\omega^2)\right],\\
&&b_0=-\frac{\epsilon^2}{8(-1-2A_{02}+A_{11}^2-2A_{20}-4A_{02}A_{20})}\left[ 
-3A_{03}+3A_{11}A_{12}-18A_{03}A_{20}
\nonumber\right.\\&&\left.
+12A_{11}A_{12}A_{20}-36A_{03}A_{20}^2
+12A_{11}A_{12}A_{20}^2-24A_{03}A_{20}^3-3A_{11}^2A_{21}-6A_{11}^2A_{20}A_{21}
\nonumber\right.\\&&\left.
+3A_{11}^3A_{30}+(-6A_{03}+4A_{11}A_{12}-24A_{03}A_{20}+8A_{11}A_{12}A_{20}-24A_{03}A_{20}^2-4A_{21}
\nonumber\right.\\&&\left.
-2A_{11}^2A_{21}-8A_{20}A_{21}+12A_{11}A_{30})\omega^2
+(-3A_{03}+A_{11}A_{12}-6A_{03}A_{20})\omega^4\right],
\end{eqnarray}
Again equating the coefficients of $\cos 2\omega t$ and $\sin 2\omega t$ from both sides of Eqs. (\ref{eq:3eq68}) and (\ref{eq:3eq69}) we get a system of four equations with unknown $a_2, \ b_2, \ a_{-2},$ and $b_{-2}$ and solving them, we get  
\begin{eqnarray}
&&a_{2}=\frac{\epsilon^2}{8\omega^2 d}\left[\{-(2+4A_{02}+A_{11}^2+4A_{20}+8A_{02}A_{20})A_{21}+3(1+2A_{02})A_{11}A_{30}\}
\nonumber\right.\\&&\left.
-2A_{11}A_{21}(-5+2A_{02}+8A_{20})
-12A_{30}(1+2A_{02}-A_{11}^2)\omega^2-8(A_{11}A_{21}+6A_{30})\omega^4
\nonumber\right.\\&&\left.
-3A_{03}A_{11}(1+2A_{20}+\omega^2)^2+A_{12}(1+2A_{20}+\omega^2)\times
\nonumber\right.\\&&\left.
\{1+2A_{02}+2A_{11}^2+2A_{20}+4A_{02}A_{20}+(-11+2A_{02}
+8A_{20})\omega^2+4\omega^4\}
\right],\\
&&b_{2}=\frac{3\epsilon^2}{8\omega^2 d}\left[A_{11}(1+2A_{20})(-A_{12}-2A_{12}A_{20}+A_{11}A_{21})
\nonumber\right.\\&&\left.
-A_{11}^3A_{30}+2\{2A_{21}+A_{11}^2A_{21}+4A_{20}A_{21}-2A_{11}(A_{12}
+2A_{12}A_{20}+3A_{30})\}\omega^2-3A_{11}A_{12}\omega^4
\nonumber\right.\\&&\left.
+A_{03}(1+2A_{20}+\omega^2)^2(1+2A_{20}+4\omega^2)
\right],\\
&&a_{-2}=\frac{\epsilon^2}{2\omega d}\left[A_{21}+2A_{02}A_{21}+2A_{11}^2A_{21}+3A_{03}(1+2A_{20}
+\omega^2)^2+A_{21}(2A_{20}+\omega^2)\times
\nonumber\right.\\&&\left.
(1+2A_{02}+4\omega^2)-3A_{11}\{A_{12}+2A_{12}A_{20}+A_{30}+2A_{02}A_{30}+(A_{12}+4A_{30})\omega^2\}
\right],\\
&&b_{-2}=\frac{3\epsilon^2\omega}{2d}[-A_{11}A_{21}+4A_{30}+A_{12}(1+2A_{20}+\omega^2)],
\end{eqnarray} 
where 
\begin{eqnarray}
&&d=1+2A_{02}+2A_{20}-A_{11}^2+4A_{02}A_{20}
+8(-1+A_{02}+A_{20})\omega^2+16\omega^4.
\end{eqnarray}
\subsection{Third order Fourier coefficients}
In a similar way using the expressions of first order coefficients $a_1,\ b_1, \ a_{-1}$, \ $b_{-1}$ and second order coefficients $a_0,\ b_0, a_2,\ b_2,\ a_{-2},\ b_{-2}$, it is possible to find the coefficients of the third order in the coefficient scheme corresponding to $\cos 3\omega t$ and $\sin 3\omega t$ and $\xi^2, \ \xi \eta,\ \eta^2, \ \xi^3, \ \xi^2\eta, \ \xi\eta^2$ and $\eta^3$. Inserting these values in Eqs. (\ref{eq:3eq68}) and (\ref{eq:3eq69}), we can compute the coefficients
of $\cos 3\omega t$ and $\sin 3\omega t$. Equating these coefficients to zero, we obtain four equations for the determination of third order Fourier coefficients $a_3,\ b_3, \ a_{-3}$ and $b_{-3}$. Solving these, we obtain 
\begin{eqnarray}
&&a_{3}=\frac{a_{3a}}{6A_{11}^2\omega+6\omega\{36\omega^2-(1+2A_{02}+9\omega^2)(1+2A_{20}+9\omega^2)\}},\\
&&b_{3}=\frac{b_{3a}}{4\{-A_{11}^2-36\omega^2+(1+2A_{02}+9\omega^2)(1+2A_{20}+9\omega^2)\}},\\
&&a_{-3}=\frac{a_{-3a}}{4\{1-A_{11}^2+2A_{20}-18\omega^2+18A_{20}\omega^2
+81\omega^4+2A_{02}(1+2A_{20}+9\omega^2)\}},\\
&&b_{-3}=\frac{b_{-3a}}{4\{1-A_{11}^2+2A_{20}-18\omega^2+18A_{20}\omega^2
+81\omega^4+2A_{02}(1+2A_{20}+9\omega^2)\}},
\end{eqnarray}
where the values of $a_{3a},\ b_{3a}, \ a_{-3a}$ and $b_{-3a}$ are given in Appendix A.
\begin{figure*}
\centering
\begin{minipage}{0.50\columnwidth}
  \centering
  \includegraphics[width=1\columnwidth]{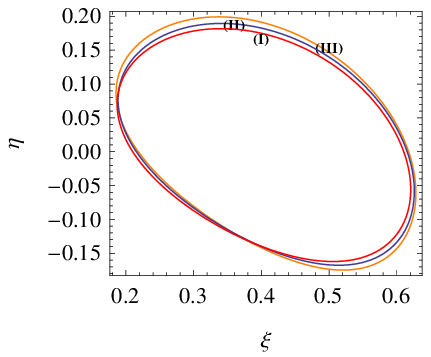}
 \caption{Periodic orbits around $L_4$ in Kepler-16 system: (I) is for $q_1=0.99, q_2=0.88$, (II) is for $q_1=0.88, q_2=0.77$, (III) is for $q_1=0.77, q_{2}=0.66$ with the orbit constant $\epsilon=0.2$. \label{fig:3fig16a}}
\end{minipage}~~~~~~~~~~~~
\begin{minipage}{.50\columnwidth}
  \centering
  \includegraphics[width=0.9\columnwidth]{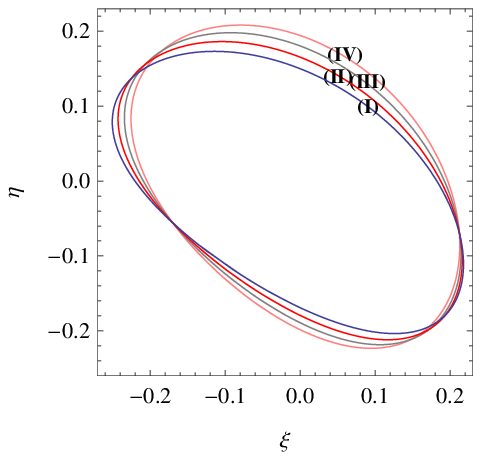}
 \caption{Periodic orbits around $L_4$ in Sun-Earth system with $q_2=1$: (I) is for $q_1=0.90$, (II) is for $q_1=0.80$, (III) is for $q_1=0.70$, (IV) is for $q_1=0.60$ with the orbit constant $\epsilon=0.2$. \label{fig:3fig15a}}
\end{minipage}
\end{figure*}
 
\begin{figure*}
\centering
\begin{minipage}{0.50\columnwidth}
  \centering
  \includegraphics[width=1\columnwidth]{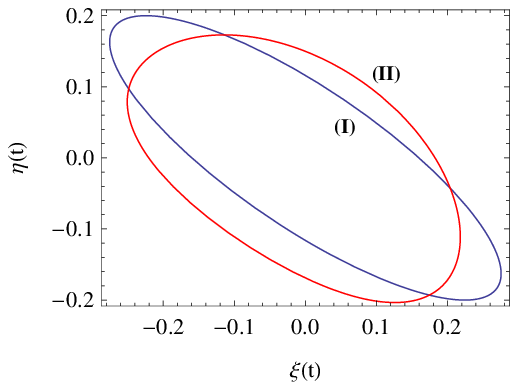}
  \caption{Periodic orbits around $L_4$ in Sun-Earth system: curve (I) is obtained using Fourier series, curve (II) is obtained by the method as in \cite{Broucke1968port.book.....B}. \label{fig:3fig19a}}
\end{minipage}~~~~~~~~~~~~
\begin{minipage}{.50\columnwidth}
  \centering
  \includegraphics[width=0.9\columnwidth]{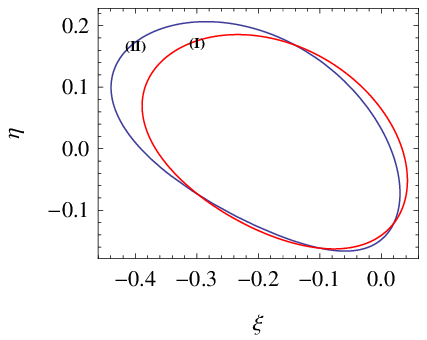}
  \caption{Periodic orbits around $L_4$ in Kepler-16 system: (I) is for $\omega=0.9$, (II) is for $\omega=0.6$,  with the orbit constant $\epsilon=0.2$. \label{fig:3fig17a}}
\end{minipage}
\end{figure*} 
 \begin{figure*}
\centering
\begin{minipage}{0.50\columnwidth}
  \centering
  \includegraphics[width=0.9\columnwidth]{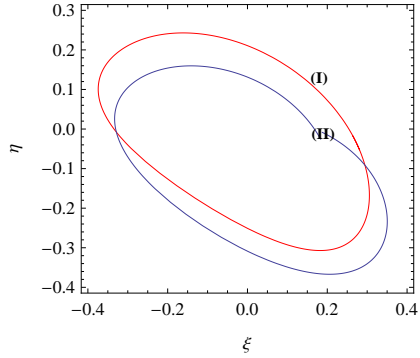}
  \caption{Comparison between Fourier-series solution and numerical solution: curve (I) is periodic orbits around $L_4$ obtained using Fourier-series method while curve (II) by RK4 numerical method in Sun-Earth system when $q_1=0.90, q_2=1$. \label{fig:3fig19arev}}
\end{minipage}~~~~~~~~~~~~
\end{figure*} 
 \begin{figure*}
 \begin{center}
 \includegraphics[width=0.70\textwidth]{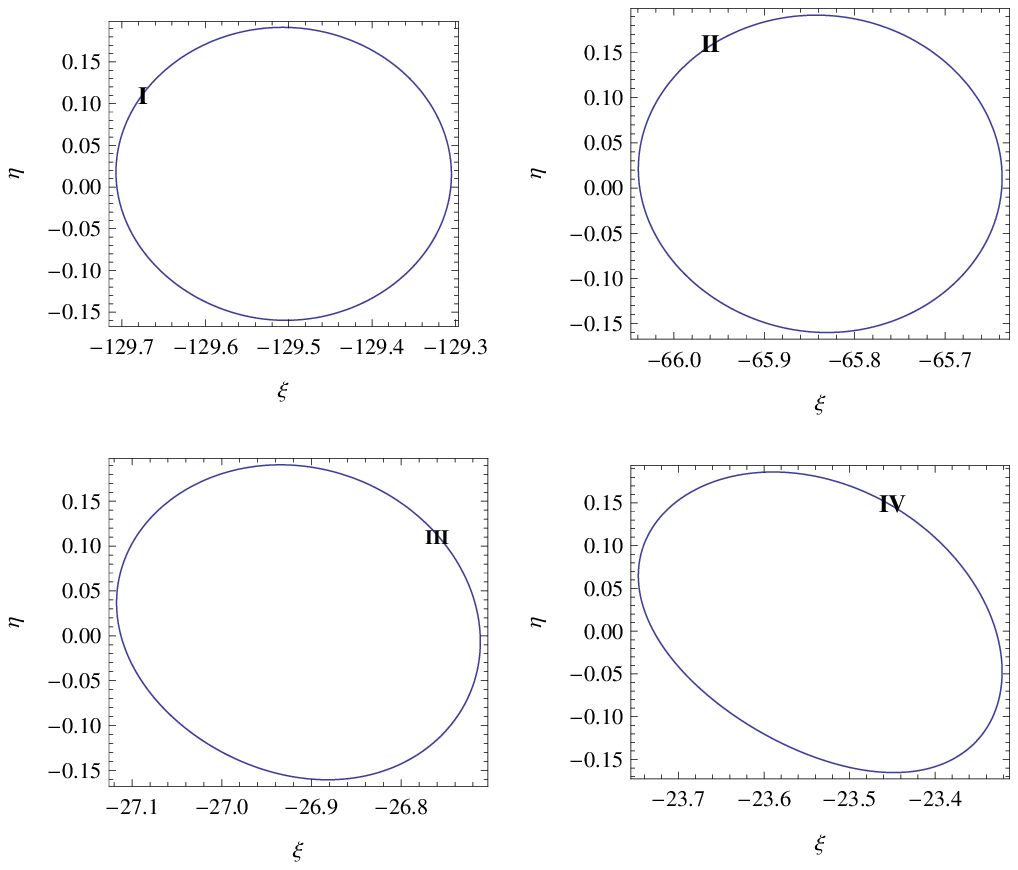}
\caption{Periodic orbits around $L_4$ of binary systems with orbit constant $\epsilon=0.2$: (I) Kepler-34, (II) Kepler-35, (III) Kepler-413, (IV) Kepler-16 . \label{fig:3fig18a}}
 \end{center}
 \end{figure*}
In Fig. \ref{fig:3fig16a} we show the periodic orbits around $L_4$ in binary system Kepler-16 for different values of parameter $q_1$ and $q_2$. We observe that for decreasing values of $q_1$ and $q_2$, the periodic orbits are expanding.
Fig. \ref{fig:3fig15a} represents the periodic orbits around $L_4$ points in Sun-Earth system for different values of radiation parameter $q_1$ when $q_2=1$ (as bigger primary Sun is only radiating). In this figure as we decrease the values of $q_1$, periodic orbits are shifting towards the origin and expanding.
A comparison of periodic orbits by Fourier-series method and as in \cite{Broucke1968port.book.....B} are shown  
in Fig. \ref{fig:3fig19a}. We have taken Sun-Earth system with $\mu=0.000003$ and $q_1=0.90, \ q_2=1$. In this case critical value of $\mu$ is $\mu_c=0.00349086$ and the four roots are pure imaginary and the roots are $\lambda_{1,2}=\pm \alpha i=\pm 0.836621i$ and $\lambda_{3,4}=\pm \beta i=\pm 0.632506i$. Hence the motion of the infinitesimal mass  around $L_4$ is stable and the general solution is of the form 
\begin{eqnarray}
&&\xi=A_1e^{i\alpha t}+A_2e^{-i\alpha t}+A_3e^{i\beta t}+A_4e^{-i\beta t},\nonumber\\\label{eq:3eq92}
&&\eta=B_1e^{i\alpha t}+B_2e^{-i\alpha t}+B_3e^{i\beta t}+B_4e^{-i\beta t}.
\end{eqnarray}
Eq. (\ref{eq:3eq92}) is composed form of two periodic motions known as long and short periodic motions with periods $T_1=\frac{2\pi}{\alpha}$ and $T_2=\frac{2\pi}{\beta}$.

In Fig. \ref{fig:3fig19arev}, a comparison is made between Fourier series solution and solution obtained by numerical methods where curve (I) represents the periodic orbits around $L_4$ by Fourier series method and curve (II) by RK4 numerical method. In this figure for Fourier series solution we have retained upto third order terms in the Fourier series and for numerical solutions we integrate the whole equations of motion. Since for Fourier series solution, we have obtained the governing equations of motion of the infinitesimal mass in the vicinity of triangular points upto third order terms neglecting the higher order terms, the discrepancies between the Fourier series solutions and numerical solutions are expected (Fig. \ref{fig:3fig19arev}). 
In Fig. \ref{fig:3fig17a} we show the periodic obits around $L_4$ point in binary system Kepler-16 for the values of $\omega=0.9$ and $\omega=0.6$. It is observed that the orbit is enlarged furthest to the left when values of $\omega$ decreases from $0.9$ to $0.6$. Further, the periodic orbits around 
$L_4$ point for four binary systems are shown in Fig. \ref{fig:3fig18a} using the actual values of radiation parameters for both 
the primaries.

 \section{Conclusions}\label{sec:3sec8}
\label{sec:con}
We have studied the motion of a infinitesimal mass in the context of the binary stellar systems Kepler-34, Kepler-35, Kepler-413 and Kepler-16. 
We have applied restricted three body problem as the model, considering gravitational and radiation effects on the particle from both 
the stars. With the help of the perturbation technique, semi-analytical expressions for the location of collinear points have been obtained.
A comparison is presented in Table \ref{tab:2a} between the analytical and numerical solution of collinear points which shows an excellent level of agreement.
We examined the linear stability of triangular points by obtaining the expressions for critical mass. We have found that critical mass depends on the radiation of both primaries. It is observed that for the 
stellar binary systems the roots of the characteristic equations are complex conjugate with one root has positive real part. Hence the motion around triangular points are unstable. 
Further, we have obtained Fourier expansions of the periodic orbits around triangular points in the CR3BP with radiation pressure from binaries.
We observed that the periodic orbits are expanding for decreasing values of $q_1$ and $q_2$. We have also observed in Sun-Earth system that as we decrease the values of $q_1$, periodic orbits are shifting towards the origin and expanding. Moreover, periodic orbits obtained by Fourier series method have been compared with that of RK4 numerical methods.
Also using the actual values of radiation parameters, we have computed periodic orbits in four binary systems.
Moreover, since detection of binary systems are increasing in number, studies of such systems provide an important contribution for future observations.

Furthermore, as the higher order approximation produce the closer orbit, the work would be extended by considering higher order terms in the governing equations of motion and in the Fourier-series. 
\begin{acknowledgements}
We are thankful to Inter-University Centre for Astronomy and Astrophysics (IUCAA), Pune, India for supporting library visits
and for the use of computing facilities.
\end{acknowledgements}
\section*{Appendix}
\begin{eqnarray}
&&A_{00}=\frac{(1-\mu)q_1}{\sqrt{c_1}}+\frac{\mu q_2}{\sqrt{c_2}}, \quad
A_{10}=-\frac{h(1-\mu)q_1}{2c_1^{\frac{3}{2}}}-\frac{(-2+h)\mu q_2}{2c_2^{\frac{3}{2}}},\quad
\nonumber\\&&
A_{01}=-\frac{(1-\mu)q_1 y_0}{c_1^{\frac{3}{2}}}-\frac{\mu q_2 y_0}{c_2^{\frac{3}{2}}},\quad
A_{11}=\frac{3h(1-\mu)q_1 y_0}{2c_1^{\frac{5}{2}}}+\frac{3(-2+h)\mu q_2 y_0}{2c_2^{\frac{5}{2}}}
\nonumber\\&&
A_{20}=\frac{(-4c_1+3h^2)(1-\mu)q_1}{8c_1^{\frac{5}{2}}}+\frac{(12-4c_2-12h+3h^2)\mu q_2}{8c_2^{\frac{5}{2}}},\quad
\nonumber\\
&&A_{02}=(1-\mu)q_1\left(-\frac{1}{2c_1^{\frac{3}{2}}}+\frac{3y_0^2}{2c_1^{\frac{5}{2}}}\right)+\mu q_2\left(-\frac{1}{2c_2^{\frac{3}{2}}}+\frac{3y_0^2}{2c_2^{\frac{5}{2}}}\right),\nonumber\\\nonumber
&&A_{30}=(1-\mu)q_1\left(\frac{3h}{4c_1^{\frac{5}{2}}}-\frac{5h^3}{16c_1^{\frac{7}{2}}}\right)
+\mu q_2(-2+h)\frac{(-20+12c_2+20h-5h^2)}{16c_2^{\frac{7}{2}}},\nonumber\\
&&A_{21}=\frac{3(4c_1-5h^2)(1-\mu)q_1 y_0}{8c_1^{\frac{7}{2}}}
+\frac{3(-20+4c_2+20h-5h^2)\mu q_2 y_0}{8c_2^{\frac{7}{2}}},\nonumber\\
&&A_{12}=\frac{3(-2+h)\mu q_2(c_2-5y_0^2)}{4c_2^{\frac{7}{2}}}+(1-\mu)q_1\left(\frac{3h}{4c_1^{\frac{5}{2}}}-\frac{15h y_0^2}{4c_1^{\frac{7}{2}}}\right),\nonumber\\
&&A_{03}=(1-\mu)q_1\left(\frac{3y_0}{2c_1^{\frac{5}{2}}}-\frac{5y_0^3}{2c_1^{\frac{7}{2}}}\right)+\mu q_2\left(\frac{3y_0}{2c_2^{\frac{5}{2}}}-\frac{5y_0^3}{2c_2^{\frac{7}{2}}}\right),\nonumber\\
&&A_{40}=\frac{(48c_1^2-120c_1 h^2+35h^4)(1-\mu)q_1}{128c_1^{\frac{9}{2}}}
+\frac{(560-480c_2+48c_2^2)\mu q_2}{128c_2^{\frac{9}{2}}}\nonumber\\
&&~~~~~~~~~~~~~~~~~~~~~+\frac{(-1120h+480c_2h+840h^2-120c_2h^2-280h^3+35h^4)\mu q_2}{128c_2^{\frac{9}{2}}},\nonumber\\\nonumber
\end{eqnarray}
\begin{eqnarray}
&&A_{31}=\frac{5(-2+h)(28-12c_2-28h+7h^2)\mu q_2 y_0}{16c_2^{\frac{9}{2}}}
+(1-\mu)q_1\left(\frac{-15hy_0}{4c_1^{\frac{7}{2}}}+\frac{35h^3y_0}{16c_1^{\frac{9}{2}}}\right),\nonumber\\
&&A_{22}=\frac{3(1-\mu)q_1(4c_1^2-5c_1h^2-20c_1y_0^2+35h^2y_0^2)}{16c_1^{\frac{9}{2}}}\nonumber\\&&
+\frac{1}{8}\mu q_2\left(-\frac{4}{c_2^\frac{5}{2}}+\frac{40y_0^2}{c_2^{\frac{7}{2}}}-\frac{5(12-4c_2-12h+3h^2)(c_2-7y_0^2)}{2c_2^{\frac{9}{2}}}\right),\nonumber\\
&&A_{13}=-\frac{5(-2+h)\mu q_2(3c_2y_0-7y_0^3)}{4c_2^{\frac{9}{2}}}
+(1-\mu)q_1\left(-\frac{15hy_0}{4c_1^{\frac{7}{2}}}+\frac{35hy_0^3}{4c_1^{\frac{9}{2}}}\right),\nonumber\\
&&A_{04}=(1-\mu)q_1\left(\frac{3}{8c_1^{\frac{5}{2}}}-\frac{15y_0^2}{4c_1^{\frac{7}{2}}}+\frac{35y_0^4}{8c_1^{\frac{9}{2}}}\right)
+\mu q_2\left(\frac{3}{8c_2^{\frac{5}{2}}}-\frac{15y_
0^2}{4c_2^{\frac{7}{2}}}+\frac{35y_0^4}{8c_2^{\frac{9}{2}}}\right),\nonumber\\
&&a_{3a}=-A_{11}\left[A_{31}\epsilon^3-3A_{31}\epsilon a_{-1}^2+4A_{21}\epsilon a_{2}-12A_{03}b_{-2}b_{-1}
\nonumber\right.\\&&\left.
-3A_{13}\epsilon  b_{-1}^2-4a_{-2}(A_{21}a_{-1}+A_{12}b_{-1})-4a_{-1}(A_{12}b_{-2}+A_{22}\epsilon b_{-1})+4A_{12}\epsilon b_{2}\right]\nonumber\\&&
+2(1+2A_{02}+9\omega^2)\{2A_{40}\epsilon^3
-6A_{40}\epsilon a_{-1}^3+6A_{30}\epsilon a_{2}-2A_{12}b_{-2}b_{-1}-A_{22}\epsilon b_{-1}^2\nonumber\\&&
-2a_{-2}(3A_{30}a_{-1}+A_{21}b_{-1})-a_{-1}(2A_{21}b_{-2}+3A_{31}\epsilon b_{-1})
+2A_{21}\epsilon b_{2}\}\nonumber\\&&
-6\omega\left[4A_{21}\epsilon a_{-2}-A_{31}a_{-1}^3-2A_{22}a_{-1}^2b_{-1}+a_{-1}(3A_{31}\epsilon^2+4A_{21}a_{2}-3A_{13}b_{-1}^2+4A_{12}b_{2})
\nonumber\right.\\&&\left.
+2\{2A_{12}\epsilon b_{-2}
+b_{-1}(A_{22}\epsilon^2+2A_{12}a_{2}-2A_{04}b_{-1}^2+6A_{03}b_{2})\}
\right],\nonumber\\
&&b_{3a}=-(1+2A_{20}+9\omega^2)\{A_{31}\epsilon^3-3A_{31}\epsilon a_{-1}^2+4A_{21}\epsilon a_{2}
-12A_{03}b_{-2}b_{-1}-3A_{13}\epsilon b_{-1}^2
\nonumber\\&&
-4a_{-2}(A_{21}a_{-1}+A_{12}b_{-1})-4a_{-1}(A_{12}b_{-2}+A_{22}\epsilon b_{-1})+4A_{12}\epsilon b_{2}\}+2A_{11}\{2A_{40}\epsilon^3
\nonumber\\&&
-6A_{40}\epsilon a_{-1}^2+6A_{30}\epsilon a_{2}-2A_{12}b_{-2}b_{-1}-A_{22}\epsilon b_{-1}^2-2a_{-2}(3A_{30}a_{-1}+A_{21}b_{-1})
\nonumber\\&&
-a_{-1}(2A_{21}b_{-2}+3A_{31}\epsilon b_{-1})
+2A_{21}\epsilon b_{2}\}
-6\omega\{12A_{30}\epsilon a_{-2}-4A_{40}a_{-1}^3+4A_{21}\epsilon b_{-2}
\nonumber\\&&
+3A_{31}\epsilon^2 b_{-1}-3A_{31}a_{-1}^2b_{-1}+4A_{21}a_{2}b_{-1}-A_{13}b_{-1}^3+4A_{12}b_{-1}b_{2}
\nonumber\\&&
+2a_{-1}(6A_{40}\epsilon^2+6A_{30}a_{2}-A_{22}b_{-1}^2+2A_{21}b_{2})\},\nonumber\\
&&a_{-3a}=-6A_{31} \epsilon ^3 \omega +\{-A_{11} A_{31}+4 A_{40} \left(1+2 A_{02}+9 \omega ^2\right)\}
a_{-1}^3-24A_{21} \epsilon  \omega  a_2
\nonumber\\&&
+4 A_{11} A_{12} \epsilon  b_{-2}-4 A_{21} \epsilon  b_{-2}
-8 A_{02} A_{21} \epsilon  b_{-2}-36 A_{21} \epsilon  \omega^2 b_{-2}+2 A_{11} A_{22} \epsilon ^2 b_{-1}
-3 A_{31} \epsilon ^2 b_{-1}\nonumber\\&&
-6 A_{02} A_{31} \epsilon ^2 b_{-1}-27 A_{31} \epsilon ^2 \omega ^2 b_{-1}+4 A_{11} A_{12}a_2 b_{-1}-4 A_{21} a_2 b_{-1}-8 A_{02} A_{21} a_2 b_{-1}
\nonumber\\&&
-36 A_{21} \omega ^2 a_2 b_{-1}+72 A_{03} \omega  b_{-2} b_{-1}+18 A_{13} \epsilon  \omega  b_{-1}^2-4A_{04} A_{11} b_{-1}^3+A_{13} b_{-1}^3\nonumber\\&&
+2 A_{02} A_{13} b_{-1}^3+9 A_{13} \omega ^2 b_{-1}^3
+4 a_{-2} \left[\epsilon \{A_{11} A_{21}-3 A_{30} \left(1+2 {A02}+9 \omega ^2\right)\}+6 A_{21} \omega  a_{-1}
\nonumber\right.\\&&\left.
+6A_{12} \omega  b_{-1}\right]+a_{-1}^2 \left[18 A_{31} \epsilon  \omega +\{-2 A_{11} A_{22}
+3 A_{31} \left(1+2 A_{02}+9 \omega ^2\right)\} b_{-1}\right]-24A_{12} \epsilon  \omega  b_2
\nonumber\\&&
+12 A_{03} A_{11} b_{-1} b_2-4 A_{12} b_{-1} b_2-8 A_{02} A_{12} b_{-1} b_2
-36 A_{12} \omega ^2 b_{-1} b_2+a_{-1} \left[3 A_{11} A_{31} \epsilon ^2
\nonumber\right.\\&&\left.
-12 A_{40} \epsilon ^2-24 A_{02} A_{40} \epsilon ^2-108 A_{40} \epsilon ^2 \omega^2+4\{A_{11} A_{21}
-3 A_{30} \left(1+2 A_{02}+9 \omega ^2\right)\} a_2
\nonumber\right.\\&&\left.
+24 A_{12} \omega  b_{-2}+24 A_{22} \epsilon\omega  b_{-1}-3 A_{11} A_{13} b_{-1}^2+2 A_{22} b_{-1}^2+4 A_{02} A_{22} b_{-1}^2+18 A_{22} \omega ^2 b_{-1}^2
\nonumber\right.\\&&\left.
+4 A_{11} A_{12}b_2-4 A_{21} b_2-8 A_{02} A_{21} b_2-36 A_{21} \omega ^2 b_2\right],\nonumber
\end{eqnarray}
\begin{eqnarray}
&&b_{-3a}=24A_{40}\epsilon^3\omega+(A_{31}+2A_{20}A_{31}-4A_{11}A_{40}
+9A_{31}\omega^2)a_{-1}^3+72A_{30}\epsilon\omega a_{2}-4A_{12}\epsilon b_{-2}
\nonumber\\&&
-8A_{12}A_{20}\epsilon b_{-2}+4A_{11}A_{21}\epsilon b_{-2}-36A_{12}\epsilon\omega^2 b_{-2}-2A_{22}\epsilon^2 b_{-1}
-4A_{20}A_{22}\epsilon^2b_{-1}
\nonumber\\&&
+3A_{11}A_{31}\epsilon^2 b_{-1}-18A_{22}\epsilon^2\omega^2 b_{-1}-4A_{12}a_{2}b_{-1}-8A_{12}A_{20}a_{2}b_{-1}+4A_{11}A_{21}a_{2}b_{-1}
\nonumber\\&&
-36A_{12}\omega^2a_{2}b_{-1}-24A_{12}\omega b_{-2}b_{-1}-12A_{22}\epsilon\omega b_{-1}^2
+4A_{04}b_{-1}^3-A_{11}A_{13}b_{-1}^3
\nonumber\\&&
+8A_{04}A_{20}b_{-1}^3+36A_{04}\omega^2 b_{-1}^3
-4a_{-2}\{\epsilon(A_{21}+2A_{20}A_{21}-3A_{11}A_{30}+9A_{21}\omega^2)
\nonumber\\&&
+18A_{30}\omega a_{-1}+6A_{21}\omega b_{-1}\}+a_{-1}^2\times
\left[-72A_{40}\epsilon\omega+\{-3A_{11}A_{31}+2A_{22}(1+2A_{20}+9\omega^2)\}b_{-1}
\right]\nonumber\\&&
+24A_{21}\epsilon\omega b_{2}-12A_{03}b_{-1}b_{2}+4A_{11}A_{12}b_{-1}b_{2}
-24A_{03}A_{20}b_{-1}b_{2}-108A_{03}\omega^2b_{-1}b_{2}-a_{-1}\{3A_{31}\epsilon^2
\nonumber\\&&
+27A_{31}\epsilon^2\omega^2+4(A_{21}+2A_{20}A_{21}-3A_{11}A_{30}+9A_{21}\omega^2)a_{2}
+24A_{21}\omega b_{-2}+36A_{31}\epsilon\omega b_{-1}
\nonumber\\&&
-3A_{13}b_{-1}^2-6A_{13}A_{20}b_{-1}^2+2A_{11}A_{22}b_{-1}^2-27A_{13}\omega^2b_{-1}^2+4A_{12}b_{2}+8A_{12}A_{20}b_{2}
\nonumber\\&&
-4A_{11}A_{21}b_{2}+36A_{12}\omega^2b_2+6A_{20}A_{31}\epsilon^2-12A_{11}A_{40}\epsilon^2\}.\nonumber
\end{eqnarray}

\bibliographystyle{spbasic}      

%
%

\end{document}